\documentclass[epj,nopacs]{svjour}

\usepackage{amsmath}
\usepackage{amssymb}
\usepackage{graphicx}
\usepackage{hyperref}
\usepackage{units}
\usepackage[switch]{lineno}

\begin{document}
\sloppy

\title{
    New limits on the resonant absorption of solar axions obtained with a $^\mathbf{169}$Tm-containing cryogenic detector
}

\author{
    A.~H.~Abdelhameed\inst{1} \and
    S.~V.~Bakhlanov\inst{2} \and
    P.~Bauer\inst{1} \and
    A.~Bento\inst{1},\inst{7} \and
    E.~Bertoldo\inst{1} \and
    L.~Canonica\inst{1} \and
    A.~V.~Derbin\inst{2} \and
    I.~S.~Drachnev\inst{2} \and
    N.~Ferreiro~Iachellini\inst{1} \and
    D.~Fuchs\inst{1} \and
    D.~Hauff\inst{1} \and
    M.~Laubenstein\inst{3} \and
    D.~A.~Lis\inst{4} \and
    I.~S.~Lomskaya\inst{2} \and
    M.~Mancuso\inst{1} \and
    V.~N.~Muratova\inst{2} \and
    S.~Nagorny\inst{5} \and
    S.~Nisi\inst{3} \and
    F.~Petricca\inst{1} \and
    F.~Proebst\inst{1} \and
    J.~Rothe\inst{1} \and
    V.~V.~Ryabchenkov\inst{6} \and
    S.~E.~Sarkisov\inst{6} \and
    D.~A.~Semenov\inst{2} \and
    K.~A.~Subbotin\inst{4} \and
    M~.V.~Trushin\inst{2} \and
    E.~V.~Unzhakov\inst{2} \and
    E.~V.~Zharikov\inst{4}
}                     
%
\mail{unzhakov\_ev@pnpi.nrcki.ru}
\institute{
    Max-Planck-Institut f{\"u}r Physik, D-80805 M{\"u}nchen, Germany \and
    NRC Kurchatov Institute, Petersburg Nuclear Physics Institute, 188309 Gatchina, Russia \and
    INFN, Laboratori Nazionali del Gran Sasso, 67010 Assergi, Italy \and
    Prokhorov General Physics Institute of the Russian Academy of Sciences, 119991 Moscow, Russia \and
    Queen's University, Physics Department, K7L 3N6 Kingston, Canada \and
    NRC Kurchatov Institute, 123182 Moscow, Russia \and
    also at: Departamento de Fisica, Universidade de Coimbra, P3004 516 Coimbra, Portugal
}
\date{Received: date / Revised version: date}
%
\newcommand{\fA}{$f_A$}							
\newcommand{\mA}{$m_A$}							
\newcommand{\EA}{$E_A$}							
\newcommand{\PhiA}{$\Phi_A$}						
\newcommand{\gAN}{$g_{AN}$}						
\newcommand{\gAg}{$g_{A\gamma}$}					
\newcommand{\gAe}{$g_{Ae}$}						
\newcommand{\omegaRatio}{$\omega_A/\omega_\gamma$}			
\newcommand{\piZero}{$\pi^0$}						
\newcommand{\fPi}{$f_\pi$}						
\newcommand{\mPi}{$m_\pi$}						
\newcommand{\Tm}{$^{169}\mathrm{Tm}$}					
\newcommand{\Kr}{$^{83}\mathrm{Kr}$}					
\newcommand{\Fe}{$^{57}\mathrm{Fe}$}					
\newcommand{\Li}{$^{7}\mathrm{Li}$}					
\newcommand{\FeSrc}{$^{55}\mathrm{Fe}$}					
\newcommand{\CoSrc}{$^{57}\mathrm{Co}$}					
\newcommand{\TmAlO}{$\mathrm{Tm}_{3} \mathrm{Al}_{5} \mathrm{O}_{12}$}	
\newcommand{\DCM}{$\mathrm{CH}_{2} \mathrm{Cl}_{2}$}			
\hyphenation{brems-strah-lung}
\abstract{
A search for resonant absorption of solar axions by {\Tm} nuclei was carried out.
A newly developed approach involving low-background cryogenic bolometer based on {\TmAlO} crystal was used that allowed for significant improvement of sensitivity in comparison with previous {\Tm} based experiments.
The measurements performed with \unit[$8.18$]{g} crystal during \unit[$6.6$]{days} exposure yielded the following limits on axion couplings: $|g_{A\gamma} (g_{AN}^0 + g_{AN}^3) \leq \unit[1.44 \times 10^{-14}]{GeV^{-1}}$ and $|g_{Ae} (g_{AN}^0 + g_{AN}^3) \leq 2.81 \times 10^{-16}$.
\PACS{
      {PACS-key}{describing text of that key}   \and
      {PACS-key}{describing text of that key}
     } 
} 
\authorrunning{A.~H.~Abdelhameed \and A.~Bento \and E.~Bertoldo et al.}
\titlerunning{New limit on the resonant absorption of solar axions\ldots}
\maketitle

\section{Introduction}\label{sec:intro}

Originally, axions were introduced as hypothetical bosons produced by a spontaneous breaking of newly introduced chiral symmetry at some energy scale {\fA}~\cite{Peccei1977,Weinberg1978,Wilczek1978}.
The initial model of ``standard'' axion assumed the symmetry-breaking scale {\fA} to be similar to the electro-weak interactions scale, but subsequently it was excluded by a series of experiments (the extensive list can be found in corresponding section of~\cite{PDG2018}).
Afterwards, the initial axion model has been expanded into two classes of ``invisible'' axion models: hadronic (KSVZ) axion~\cite{Kim1979,Shifman1980} and GUT (DFSZ) axion~\cite{Dine1981,Zhitnitski1980}.
These models allow {\fA} to be arbitrary large, therefore reducing the expected axion mass and suppressing the axion interactions with ordinary matter, effectively rendering it ``invisible''.

The limit on the axion mass {\mA} is obtained as a consequence of the experimental limits on the effective coupling constants of axion with photons (\gAg), electrons (\gAe) and nucleons (\gAN).
Axion coupling constants appear to be significantly model-dependent, so in principle it is possible to consider more general class of axion-like particles (ALPs) with their masses and coupling constants being independent parameters.
Axions and ALPs remain suitable dark matter candidates, motivating the experimental effort to search for these particles.

The axion mass {\mA} can be expressed through the properties of \piZero-meson~\cite{Bardeen1978}:
\begin{equation}\label{eq:axion_mass}
    m_A =
        \frac{m_\pi f_\pi}{f_A}
        {\left(
            \frac{z}{(1+z)(1+z+w)}
	\right)}^{1/2},
\end{equation}
where {\mPi} and {\fPi} are respectively the pion mass and decay constant, while $z = m_u/m_d$ and $w = m_u/m_s$ are the quark mass ratios.

In a laboratory environment axions could be potentially observed via various processes with different axion couplings.
The Primakoff effect allows the conversion of axion into a detectable photon inside strong magnetic fields ({\gAg}) or the axion decay into 2 $\gamma$-quanta.
Axion interaction with electrons of atomic shells ({\gAe}) can cause axio-electric effect (similarly to the photo-ionization) or Compton-like processes.
Finally, since the axion is a pseudoscalar boson it can undergo resonant absorption or emission in nuclear transitions of magnetic type ({\gAN}).

The resonant absorption can be used for detection of solar axions in experimental setup with a target containing a nuclide with magnetic type transition to the ground state.
The general idea behind this approach is that, due to the presence of {\gAN} coupling, the axions could be resonantly absorbed by the target nucleus $N$ possessing the relevant excited state. After the absorption, the excited nucleus $N^*$ will consequently discharge, emitting the $\gamma$-quantum: $A + N \rightarrow N^* \rightarrow N + \gamma$.
The proposals for experiments aimed for registration of monochromatic solar axions produced by {\Fe}, {\Li}, {\Kr} and {\Tm} nuclei were originally made in~\cite{Moryiama1995,Krcmar2001,Jakovcic2004,Derbin2009}, correspondingly.

The main benefit of {\gAN}-based detection technique comes from the resonant nature of the absorption process, which provides high reaction cross-section and, therefore, a possibility of achieving competitive sensitivity even with a relatively small-scale experimental setup.
A ``solid target + semiconductor detector'' layout has been successfully employed for previous axion searches with various targets ({\Li}~\cite{Belli2008}, {\Tm}~\cite{Derbin2010}, {\Fe}~\cite{Derbin2011a}).
On the other hand, {\Fe} and {\Tm} nuclei have low-energy excited states with significant conversion ratios ($\eta \sim 10^{-3}$), so the most transitions would actually produce conversion or Auger electrons and characteristic X-rays, instead of nuclear $\gamma$-quanta.
Intensive self absorption of these particles inside the target material effectively limits the usable target mass by several grams, thus constraining the potentially achievable sensitivity of this approach.

A natural solution for this problem would be the introduction of the target material inside the active volume of the detector.
This approach was implemented in experiments with gaseous {\Kr} target and proportional counter located at the underground facility of Baksan Neutrino Observatory~\cite{Derbin2017,Gavrilyuk2018}.

In this paper we aim to detect solar axions via the resonant absorption by {\Tm} target, similarly to a series of previous axion searches with {\Tm} targets performed at Petersburg Nuclear Physics Institute~\cite{Derbin2009,Derbin2010,Derbin2011}.
The measurement presented here uses the recently developed approach with cryogenic bolometer based on Tm-containing crystal of a garnet family (\TmAlO)~\cite{Bertoldo2020}.
We show a significant improvement of the experimental sensitivity thanks to the inclusion of {\Tm} inside the active volume of the detector and demonstrate the potential feasibility of this approach for a kg-scale installation.

\section{Axion rate estimation}\label{sec:solar_axions}

The most intense source of axions for an experiment based on Earth is constituted by the Sun.
There are several expected mechanisms of axion production that can take place inside stars.
Axions can be produced as a result of Primakoff effect due to the axion-photon coupling ({\gAg}).
The axion-electron coupling ({\gAe}) allows for several axion-yielding reactions: atomic de-excitation and recombination, electron-nucleus and electron-electron bremsstrahlung and Compton-like scattering.
Finally, stellar cores possess high enough temperatures for the thermal excitation of low-energy nuclear levels of magnetic type ($\sim1$~keV scale), which could emit axions during the de-excitation ({\gAN} coupling).

\subsection{Solar axion flux}\label{subs:solar_flux}

The axion-photon coupling is determined by the following expression~\cite{Kaplan1985,Srednicki1985}:
\begin{equation}\label{eq:gAg}
    g_{A\gamma} =
        \frac{\alpha}{2\pi f_A}
        \left(
            \frac{E}{N} -
            \frac{2(4 + z)}{3(1 + z)}
        \right) =
        \frac{\alpha}{2\pi f_A} C_{A\gamma}
\end{equation}
where $\alpha \approx 1/137$ is the fine structure constant and $E/N$ is the ratio between electromagnetic and color anomalies.
The value of $E/N$ depends on the particular axion model: in case of DFSZ-axion $E/N = 8/3$ while in the original KSVZ model $E/N = 0$~\cite{Kaplan1985}.

\begin{figure}[t]
    \includegraphics[width=.485\textwidth]{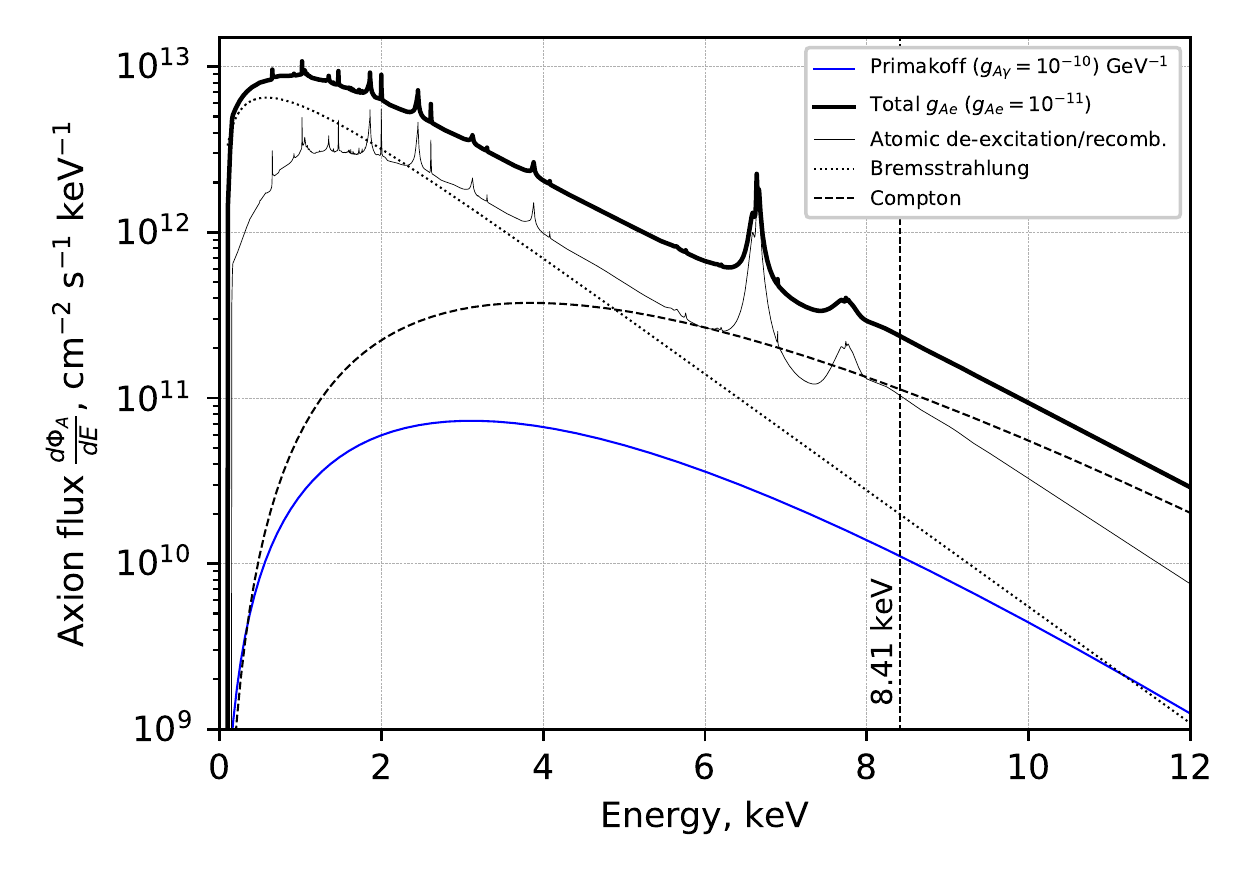}
    \caption{The calculated energy spectra of solar axions produced due to the axion-photon~\cite{Zioutas2005} and axion-electron couplings~\cite{Redondo2013}.
    The spectra are calculated in assumption of massless axion ($m_A = 0$) for nominal {\gAg} and {\gAe} values relevant to the range of experimental sensitivity.}\label{fig:solar_flux}
\end{figure}

The differential energy spectrum of Primakoff axions is calculated~\cite{vanBibber1989,Creswick1998,Zioutas2005} based on the radial distributions of temperature and electron density provided by the Standard Solar Model (SSM).
The shape of Primakoff axion spectrum calculated for nominal value $g_{A\gamma} = \unit[10^{-10}]{GeV^{-1}}$ is presented in Figure~\ref{fig:solar_flux}.
The continuous flux has a maximum at \unit[$\sim4$]{keV} and becomes negligible at energies beyond \unit[$20$]{keV}.
In case of {\Tm} as a target, the axion flux at \unit[$8.41$]{keV} would remain relatively significant at about \unit[$\sim 10$]{\%} of its maximum value.

The axion-electron coupling depends significantly on the type of axion model.
In case of DFSZ axion the direct coupling to leptons is allowed and the constant {\gAe} depends on electron mass $m_e$ as:
\begin{equation}\label{eq:gAe}
    g_{Ae} = \frac{1}{3} \cos^2 \beta \cdot \frac{m_e}{f_A}
\end{equation}
where $\beta$ is an arbitrary angle.

In KSVZ model axions can not interact with leptons directly, but the coupling via the radiative loops remains possible~\cite{Srednicki1985}:
\begin{equation}\label{eq:gAe_ksvz}
    g_{Ae} = \frac{3\alpha^2 n m_e}{2 \pi f_A}
        \left(
            \frac{E}{N} \ln{\frac{f_A}{m_e}}
            - \frac{2}{3} \cdot \frac{4 + z + w}{1 + z + w}
            \ln{\frac{\Lambda}{m_e}}
        \right)
\end{equation}
where the QCD cutoff scale $\Lambda \approx \unit[1]{GeV}$.
Consequently, in this case the axion-electron coupling is suppressed by a factor $\alpha^2$.

The expected flux of axions produced in the Sun via the axion-electron coupling ({\gAe}) is calculated using cross-sections for Compton processes~\cite{Pospelov2008,Gondolo2009} and bremsstrahlung~\cite{Zhitnitskij1979}, SSM data on the electron gas density, temperature distribution and concentrations of various elements~\cite{Derbin2011,Kekez2009}.
A recent work~\cite{Redondo2013} also includes the axion production via the processes of atomic recombination and de-excitation, which adds additional linear structure on top of the continuous Compton/bremsstrahlung spectra.

The shape of total {\gAe}-related axion spectrum (together with Compton and bremsstrahlung components) calculated for the nominal value of $g_{Ae} = 10^{-11}$ is given in Figure~\ref{fig:solar_flux}.
At lower energies below \unit[$\sim5$]{keV} bremsstrahlung axions constitute the most part of {\gAe} axion flux, while above \unit[$\sim5$]{keV} Compton axions become dominant.

It is important to note that for both axion couplings the value of solar flux $\Phi_A$ at a given axion energy {$E_A$} appears to be proportional to the square of the relevant coupling constant:
\begin{equation}\label{eq:flux}
    \Phi_{Ax}({E_A}) \propto C_{Ax} \cdot g_{Ax}^2
\end{equation} 
where $C_{Ax}$ is a constant determined for a given axion coupling $g_{Ax}$.

\subsection{Resonant absorption of axions by atomic nuclei}\label{subs:absorption_rate}

The cross-section for the resonant absorption of incident solar axions with energy {\EA} is expressed in a similar fashion as the conventional $\gamma$-ray absorption, corrected by emission probability {\omegaRatio}.
The axion absorption rate for {\Tm} can be presented as~\cite{Derbin2009}:
\begin{equation}\label{eq:absorption_rate}
    R_A =
        \pi \sigma_{0\gamma}\Gamma
        \frac{d\Phi_A}{dE_A} (E_A = 8.41\ \mathrm{keV})
        \left(
            \frac{\omega_A}{\omega_\gamma}
        \right)
\end{equation}
where $\sigma_{0\gamma}$ is the maximum cross-section of $\gamma$-ray absorption, $\Gamma$ is the width of energy level, $\omega_A$ and $\omega_\gamma$ are respectively the probabilities of axion or photon emission.

The probability ratio {\omegaRatio} was calculated in~\cite{Donnelly1978,Avignone1988} using the long wave approximation:
\begin{equation}\label{eq:omega_ratio}
    \frac{\omega_A}{\omega_\gamma} =
        \frac{1}{2 \pi \alpha}
        \frac{1}{1 + \delta^2}
        {\left[
            \frac{g^0_{AN} \beta + g^3_{AN}}{(\mu_0 - 0.5)\beta + \mu_3 - \eta}
	\right]}^2
        \left(
            \frac{p_A}{p_\gamma}
        \right)^3
\end{equation}
where $p_\gamma$ and $p_A$ are respectively the photon and axion momenta; $\mu_0 = \mu_p + \mu_n \approx 0.88$ and $\mu_3 = \mu_p - \mu_n \approx 4.71$ are the isoscalar and isovector nuclear magnetic momenta, $\beta$ and $\eta$ are parameters derived from the nuclear matrix elements of a particular target isotope.
In case of {\Tm}, using the one-particle approximation, these parameters can be approximated as $\beta \approx 1.0$ and $\eta \approx 0.5$, yielding the expression for {\omegaRatio} as~\cite{Derbin2009}:
\begin{equation}\label{eq:omega_ratio_gAN}
    \frac{\omega_A}{\omega_\gamma} =
	1.03{(g^0_{AN} + g^3_{AN})}^2 {(p_A/p_\gamma)}^3
\end{equation}

In the framework of KSVZ axion model, the axion-nucleon coupling {\gAN} consisting of the isoscalar $g^0_{AN}$ and isovector $g^3_{AN}$ terms can be expressed through the {\fA} value~\cite{Kaplan1985,Srednicki1985}:
\begin{equation}\label{eq:gAN0}
    g^0_{AN} =
        - \frac{m_N}{6 f_A}
        {\left[
	    2S + (3F - D)
            \frac{1 + z + 2w}{1 + z + w}
	\right]}
\end{equation}
and
\begin{equation}\label{eq:gAN3}
    g^3_{AN} =
        - \frac{m_N}{2 f_A}
        \left[
	    (F + D)
            \frac{1 - z}{1 + z + w}
        \right]
\end{equation}
where $m_N \approx \unit[939]{MeV}$ is the nucleon mass, $z$ and $w$ are quark mass ratios, and $F$, $D$, $S$ are axial and singlet coupling parameters.
The values of $F$ and $D$ are experimentally obtained from the observations of hyperon semi-leptonic decays~\cite{Mateu2005}: $F = 0.462 \pm 0.011$ and $D = 0.808 \pm 0.006$, for various solar axion fluxes.

The singlet coupling parameter $S$ represents the contribution of quarks to the polarization of the nucleon.
The experimental restrictions on the value of $S$, obtained in~\cite{Airapetian2007,Alexakhin2007}, are $(0.27 \le S \le 0.41)$.
Nevertheless, in further calculations we assume $S = 0.5$ for convenience of result comparison, since this value is commonly used in previous works and in other experiments.

The model independent expression for the rate of axion absorption by {\Tm} nucleus (\ref{eq:absorption_rate}) then can be derived from flux expression (\ref{eq:flux}) and {\omegaRatio} ratio (\ref{eq:omega_ratio_gAN}):
\begin{equation}\label{eq:R_A_model_indep}
    R_A =
	C_{Ax} \cdot g^2_{Ax} \cdot {(g^0_{AN} + g^3_{AN})}^2 \cdot
	{(p_A/p_\gamma)}^3
\end{equation}
where counting rate $R_A$ is expressed here in $\unit{atom^{-1}} \cdot \unit{s^{-1}}$ units.
The constant $C_{Ax}$ has a cumulative value defined by axion model parameters, properties of target nucleus, etc. --- in case of {\Tm} target $C_{A\gamma} = 104$ and $C_{Ae} = 2.76 \times 10^{5}$~\cite{Derbin2009,Redondo2013}.

Then, using the relations between axion mass {\mA} and axion-nuclei coupling {\gAN}~(\ref{eq:gAN0},~\ref{eq:gAN3}) it becomes possible to express the absorption rate as a function of axion coupling $g_{Ax}$ and {\mA} (in \unit{eV} units):
\begin{equation}\label{eq:R_A_gag_ma}
    R_A =
	C^\prime_{Ax} \cdot g^2_{Ax} \cdot m_A^2 \cdot {(p_A/p_\gamma)}^3 \\
\end{equation}
\begin{equation*}
    (C^\prime_{A\gamma} = 4.08 \times 10^{-13},\ 
    C^\prime_{Ae} = 1.03 \times 10^{-9})
\end{equation*}
Finally, by employing expressions for {\mA}~(\ref{eq:axion_mass}), {\gAg}~(\ref{eq:gAg}) and {\gAe} (\ref{eq:gAe},~\ref{eq:gAe_ksvz}) we obtain the dependence of axion absorption rate $R_A$ directly on the axion mass {\mA} (in \unit{eV} units):
\begin{equation}\label{eq:R_A_ma}
R_A = C^{\prime\prime}_{Ax} \cdot m_A^4 \cdot {(p_A/p_\gamma)}^3
\end{equation}
\begin{equation}
    (C^{\prime\prime}_{A\gamma} = 6.64 \times 10^{-32},\ 
    C^{\prime\prime}_{Ae} = 8.08 \times 10^{-31}) \nonumber
\end{equation}

The total number of expected ``axion'' events is determined by the target mass (i.e.~number of {\Tm} nuclei), detector efficiency and total live time of the measurement.
The detection probability of the resulting ``axion'' peak depends on the background level and on the energy resolution of the experiment.

\section{Cryogenic bolometer and experimental setup}\label{sec:exp_setup}

In a recent work~\cite{Bertoldo2020}, we demonstrated the possibility to operate a cryogenic bolometer based on the thulium-containing crystal {\TmAlO}.
This first prototype showed promising results, but the energy threshold achieved was rather far from meeting the minimum benchmark to be sensitive to the resonant absorption of solar axions in {\Tm}.

In order to improve the energy threshold and the energy resolution of a cryogenic bolometer based on a {\TmAlO} crystal the type of phonon sensor has been changed, replacing the Neutron Transmutation Doped (NTD) sensor with a Transition Edge Sensor (TES).
First, the same {\TmAlO} crystal used in~\cite{Bertoldo2020} has been processed with dichloromethane (CH$_{2}$Cl$_{2}$) in order to remove the glue and the NTD sensor.
After that, a CRESST-like TES~\cite{Rothe2018} has been evaporated on the crystal surface.

The TES is constituted by a thin strip of tungsten with two large aluminum pads partially overlapping the tungsten layer.
These aluminum pads have two different features: they serve as phonon collectors and bond pads.
These pads are connected via a pair of \unit[$25$]{$\mu$m} aluminum bond wires through which the bias current is injected.
The tungsten film is also connected by a long and thin strip of gold to a thicker gold bond pad on which a \unit[$25$]{$\mu$m} gold wire is bonded.
This bond serves as thermal link between the sensor and the heat bath at \unit[$\sim 10$]{mK}.

On the same surface, but separated from the TES, we also evaporate a heater.
The heater is made of a thin strip of gold with two aluminum pads deposited on top.
These pads are also bonded with a pair of 25~$\mu$m aluminum bond wires through which a tunable current can be injected to maintain the TES at the desired temperature.
The heater is also used to inject artificial pulses in order to monitor the detector response over time and to refine the energy calibration during the data analysis.

\begin{figure}[t]
    \centering
    \includegraphics[width=.485\textwidth]{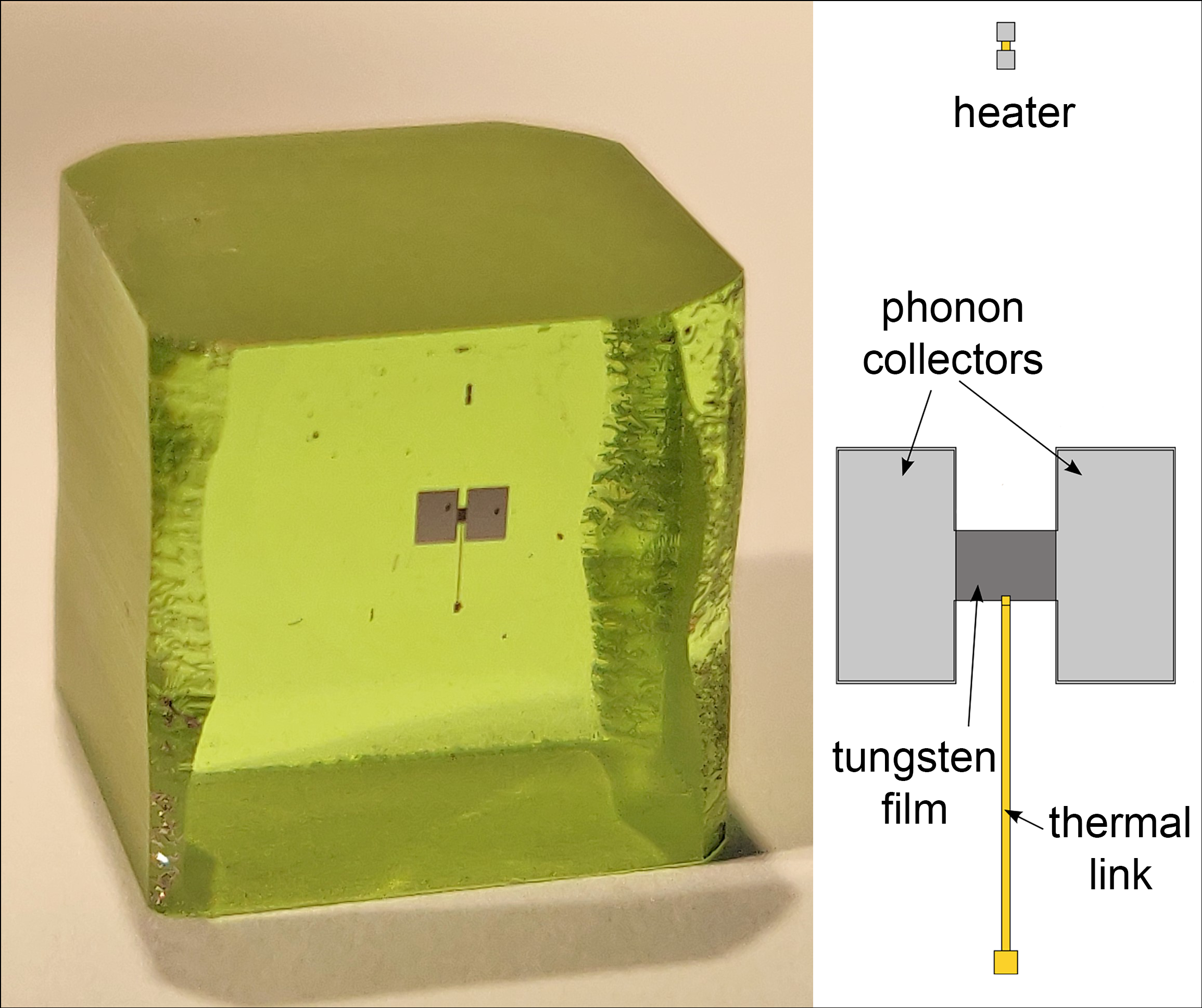}
    \caption{
       \textbf{Left}: {\TmAlO} crystal after the TES deposition.
        It is possible to see two large aluminum phonon collectors (light gray) evaporated on top of a darker strip of tungsten.
        Closer to the upper edge of the crystal surface there is the heater made of a thin strip of gold with two aluminum pads deposited on top.
	\textbf{Right}: A sketch of a similar TES design~\cite{phdanja}.
    }\label{fig:crystal_TES}
\end{figure}

We would like to highlight that this is the first time a TES is directly evaporated on a crystal containing $^{169}$Tm.
In Figure~\ref{fig:crystal_TES} we show the crystal after the TES deposition along with a sketch of the TES design.

After the TES deposition, the crystal was placed inside a copper holder where it was held in position by a pair of CuBe clamps.
Inside the holder, we place a {\FeSrc} X-ray source with activity of \unit[$\sim 0.4$]{Bq} at a distance of \unit[$\sim 1$]{mm} from one of the crystal surfaces.
This X-ray source is used for the energy calibration of the detector.
Finally, the holder is mechanically coupled to the coldest stage of a Leiden Cryogenics dilution refrigerator located in an above-ground laboratory at the Max Planck Institute for Physics (MPP) in Munich, Germany.
It has to be noted that in this laboratory there is no shielding against environmental and cosmic radiation.

A temperature of \unit[$\sim10$]{mK} has been maintained over the course of the whole run at the coldest stage of the dilution refrigerator.
The TES has a critical temperature $T_{C} = \unit[23]{mK}$, thus the operating point is stabilized around this value injecting an appropriate current through the heater.
The readout of the TES is obtained with a commercial SQUID system\footnote{Applied Physics System model 581 DC SQUID}, combined with a CRESST-like detector control system~\cite{Angloher2009}. 
The start of the run has been reserved for a first energy calibration with a {\CoSrc} source placed outside of the dilution refrigerator.
After this initial calibration, we have collected background data for solar axion search.

\section{Data analysis and results}\label{sec:analysis}

The background data acquisition has lasted for \unit[$6.60$]{days} of total measurement time.
To precisely evaluate the effective measurement time in the region of interest we created a copy of the data where we blindly inject simulated pulses of \unit[$8.41$]{keV} with a rate of \unit[1.6]{mBq$\cdot$s$^{-1}$}, $\sim 1000$ times smaller than the total rate observed from environmental radioactivity.
The data with the simulated pulses were triggered and analyzed in the same way as the background data collected, hence the fraction of survived simulated pulses corresponds to the survival probability of a hypothetical signal.After the trigger, the effective measurement time in the energy region of interest is 3.89 days, with a significant reduction to the respect of the total measurement time.
This is due to the trigger dead time, which is naturally high in an above-ground experimental setup.

One stability cut and two quality cuts are applied to the data, with the effective measurement time further reduced to \unit[$3.86$]{days}.
The stability cut rejects the periods of time when the detector is not in the desired working point, while the quality cuts reject pile-up events and artifacts in the data.
The quality cuts are based on two different pulse shape parameters.
The overall exposure is equal to \unit[$31.6$]{g$\cdot$day} with a $^{169}$Tm exposure of \unit[$19.2$]{g$\cdot$day}.

\begin{figure}[t]
    \centering
    \includegraphics[width=.485\textwidth]{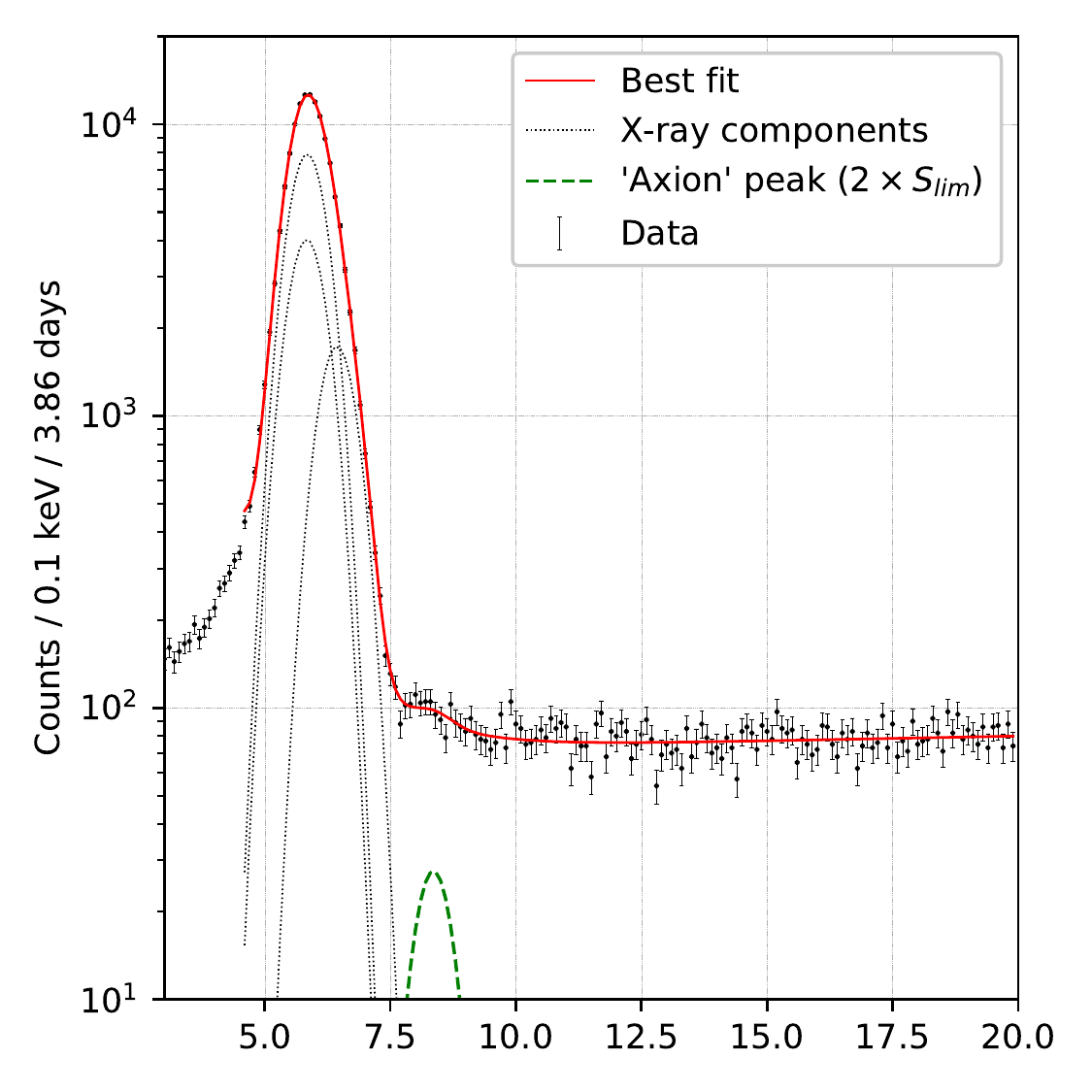}
    \caption{Spectrum of events obtained during the live time of \unit[$3.9$]{days} by {\TmAlO} bolometer in \unit[$3-20$]{keV} energy interval with \unit[$0.1$]{keV} binning.
    The result of fitting by model~(\ref{eq:fit_model}) is presented by solid line.
    The presumed ``axion'' peak with area $S = 2 \times S_{lim}$ is shown by the dashed line (the area is increased to improve visibility).
    }\label{fig:spectrum_fit}
\end{figure}

The acquired spectrum of events in \unit[$3 - 20$]{keV} energy interval is presented in Figure~\ref{fig:spectrum_fit}.
The energy calibration has been performed using a combination of the injected heater pulses and the characteristic peak induced by the presence of the {\FeSrc} source.
The injected heater pulses have three different amplitudes and are continually sent throughout the data acquisition.
Using this information we can monitor the stability of the detector and correct for any unwanted drift of the operating point.
The energy resolution of the detector is not sufficient to resolve K$_{\alpha1}$, K$_{\alpha2}$ and K$_{\beta1+3}$ characteristic X-ray lines of Mn induced by the {\FeSrc} source, thus in the spectrum only a single peak centered around \unit[$5.895$]{keV} is visible.

There is no significant excess of events in the vicinity of \unit[$8.41$]{keV}.
Hence, in order to determine the upper limit on the number of events in the ``axion peak'' a maximum likelihood method has been employed.
The fit function is chosen as a sum of exponential background, three Gaussians describing K$_{\alpha1}$, K$_{\alpha2}$, K$_\beta$ X-rays of Mn and the expected ``axion'' peak, all with the same energy resolution~$\sigma$:
\begin{equation}\label{eq:fit_model}
    N(E) = a + b \cdot E + c \cdot e^{-\frac{(E -4)}{d}}
        + \sum_{i=1}^4
	S_i e^{-\frac{{(E_i - E)}^2}{2\sigma^2}}
\end{equation}
The shape of the unresolved Mn X-ray peak is described by sum of three Gaussians, representing $K_{\alpha1}$, $K_{\alpha2}$ and $K_{\beta}$ lines.
All Gaussian positions, including the axion peak, are set relative to the position of the brightest $K_{\alpha1}$ line ($E_{\alpha1}$).
The intensities of $K_{\alpha1}$, $K_{\beta}$ and axion peak are free, while the intensity of the $K_{\alpha2}$ is set relatively to $S_{\alpha1}$.
In total, there are 9 free fit parameters: 4 background coefficients ($a, b, c, d$) and 5 peak parameters  ($\sigma$, $E_{\alpha1}$, $S_{\alpha1}$, $S_{\beta}$, $S_A$).
The exponential model of the background fails below \unit[$5$]{keV}, so in order to avoid the introduction of additional parameters we chose to raise the lower border of fit interval to \unit[$4.6$]{keV}, since this non-linear background should not make any significant contribution at \unit[$8.31$]{keV}.
The best fit with reduced chi-squared criterion $\chi^2/ N_{DoF} = 171.1/(154 - 9) = 1.18$ is presented in Fig.~\ref{fig:spectrum_fit} by a solid line.
The determined energy resolution amounted to \unit[$\sigma = 0.370$]{keV}.

In order to determine the upper limit on the \unit[$8.41$]{keV} peak intensity a standard approach of $\chi^2$-profiling was employed.
The value of $\chi^2$ is determined for different fixed values of $S_A$ while the other parameters remain unconstrained.
The obtained probability function $P(\chi^2)$ is normalized to unity for $S_A\geq 0$.
The upper limit estimated in this manner is $S_{lim}= 128$ at \unit[$90$]{\%} confidence level.

The upper limit on the amount of measured ``axion'' events $S_{lim}$ depends on the detection efficiency $\epsilon$, the number of {\Tm} nuclei $N_{Tm}$, the measurement time $T$, and axion resonant absorption rate $R_A$ calculated in Section~\ref{subs:absorption_rate}:
\begin{equation}
    \epsilon \cdot N_{Tm} \cdot T \cdot R_A \leq S_{lim}
\end{equation}
In case of {\TmAlO} crystal the detection efficiency is $\epsilon \approx 1$ since the target material is located inside the active volume of the detector.
The number of target {\Tm} nuclei in \unit[$8.18$]{g} {\TmAlO} crystal is $N_{Tm} = 1.77 \times 10^{22}$.
The exposure time left after the application of data selection cuts is equal to $T = \unit[3.86]{days}$ and $R_A$ is the axion resonant absorption rate for {\Tm} defined earlier by expressions~(\ref{eq:R_A_model_indep},~\ref{eq:R_A_gag_ma},~\ref{eq:R_A_ma}).
In accordance with these equations and in assumption that $(p_A/p_\gamma)^3 \approx 1$, which holds for axion masses below \unit[$\sim 2$]{keV}, our measurement yields the following limits on axion-photon coupling:

\begin{align}\label{eq:gag_limits}
    \left| g_{A\gamma} \cdot (g_{AN}^0 + g_{AN}^3) \right| &\leq \unit[1.44 \times 10^{-14}]{GeV^{-1}} \\
    \left| g_{A\gamma} m_A \right| &\leq 2.31 \times 10^{-7} \nonumber
\end{align}
\noindent
and axion-electron coupling:

\begin{align}\label{eq:gae_limits}
    \left| g_{Ae} \cdot (g_{AN}^0 + g_{AN}^3) \right| &\leq 2.81 \times 10^{-16}\\
    \left| g_{Ae} m_A \right| &\leq \unit[4.59 \times 10^{-9}]{eV} \nonumber
\end{align}
\noindent
The axion mass {\mA} here is expressed in \unit{eV} units and {\gAg} is expressed in \unit{GeV$^{-1}$} units, while {\gAe} and {\gAN} are dimensionless.

\begin{figure}[t]
    \centering
    \includegraphics[width=.485\textwidth]{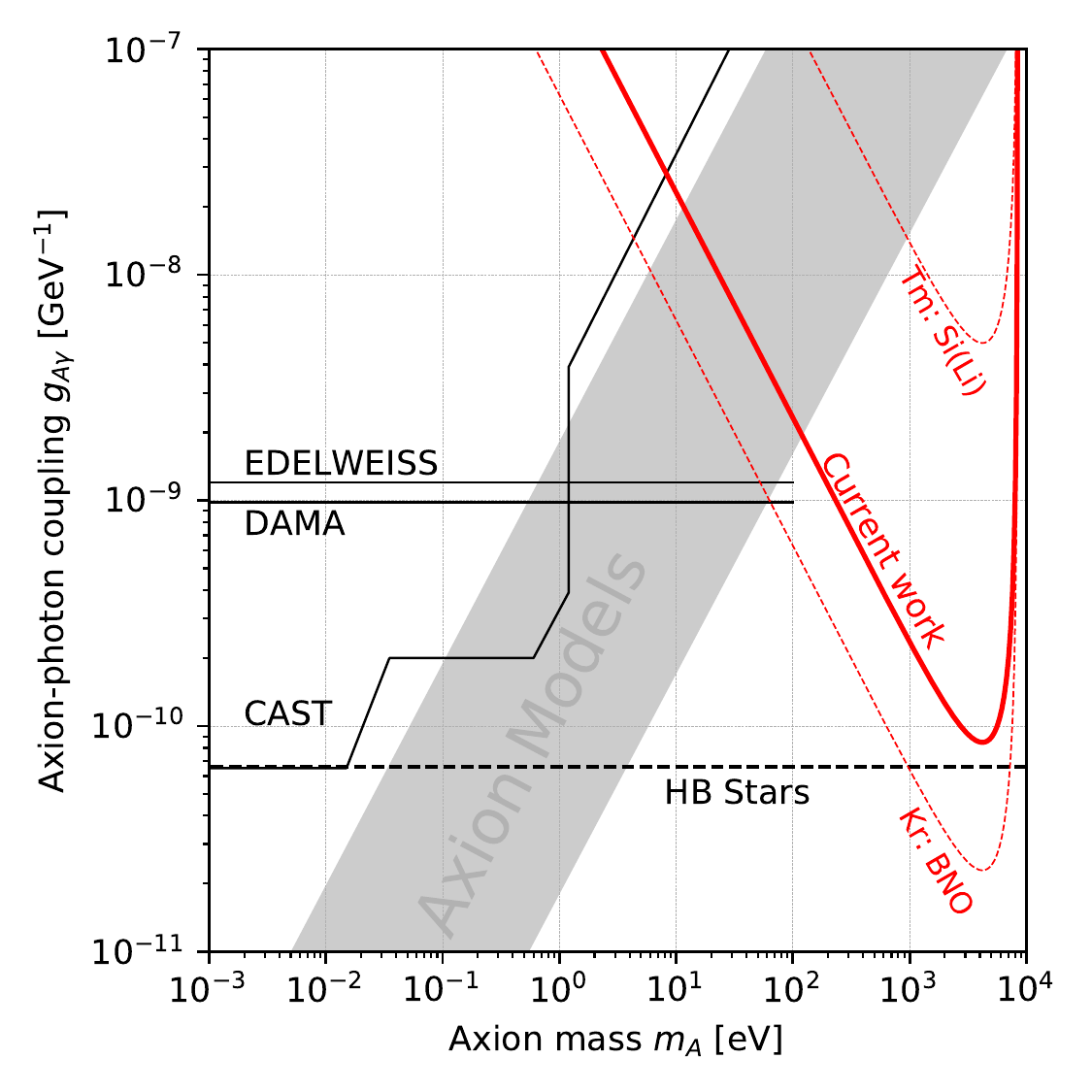}
    \caption{Axion-photon coupling {\gAg} limits obtained in current work in comparison with other experiments (DAMA~\cite{Bernabei2001}, EDELWEISS~\cite{Armengaud2013}, CAST~\cite{Anastassopoulos2017}, {\Tm}-Si(Li)~\cite{Derbin2010}, {\Kr}-gas counter~\cite{Gavrilyuk2018}) and astrophysical bounds (horizontal branch stars lifetime~\cite{Ayala2014}).}\label{fig:gag_limit}
\end{figure}

\begin{figure}[t]
    \centering
    \includegraphics[width=.485\textwidth]{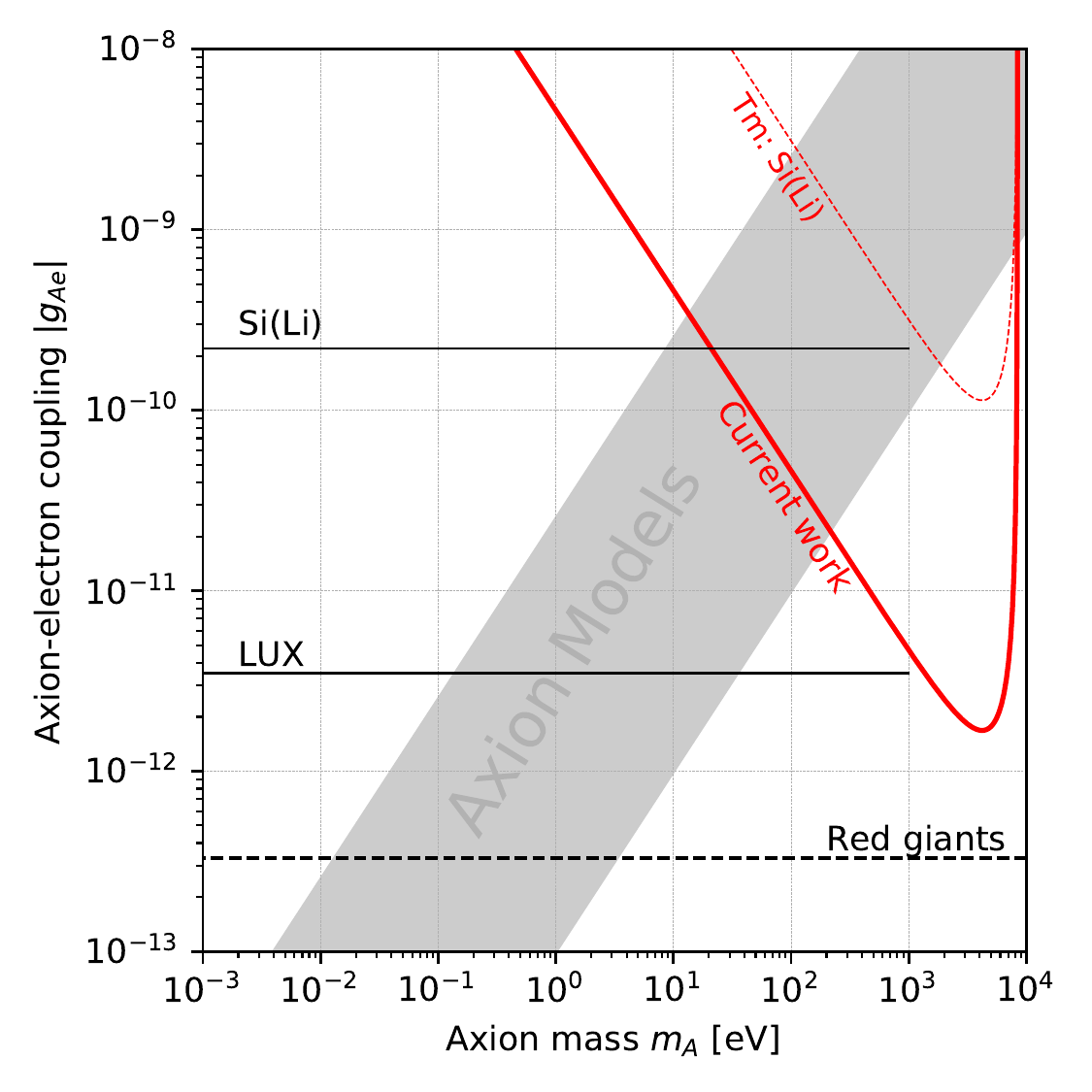}
    \caption{Axion-electron coupling {\gAe} limits obtained in current work in comparison with other experiments (axio-electric effect on Si~\cite{Derbin2012}, LUX~\cite{Akerib2017}) and astrophysical bounds (red giant cooling rate~\cite{Raffelt1995}).}\label{fig:gae_limit}
\end{figure}

The exclusion plots for the axion parameter space are given in Figures~\ref{fig:gag_limit} and~\ref{fig:gae_limit} along with comparison with other experiments and astrophysical bounds.
The limit obtained in this work significantly exceeds the best previous result achieved with {\Tm} target~\cite{Derbin2010}.
The {\gAg} limit obtained with {\Kr}~\cite{Gavrilyuk2018} still remains unsurpassed, although the current limits achieved with only \unit[$8.18$]{g} crystal and \unit[$3.86$]{days} of live time are competitive, and show potential due to the scalability of the experiment.

It should also be noted that the particular values of {\Tm} nuclear matrix elements make it a favorable axion target, since the probability ratio {\omegaRatio} never vanishes for any combination of model parameters, unlike the case of {\Fe} and {\Kr} nuclei~\cite{Derbin2011a}.

\section{Conclusions}\label{sec:conclusions}

In this work we present the first successful investigation of the resonant absorption of solar axions in {\Tm} employing a cryogenic bolometer.
The cryogenic bolometer is constituted by a \unit[$8.18$]{g} {\TmAlO} crystal with a TES directly evaporated on the crystal surface.
We have collected data for \unit[$3.86$]{days} of effective measurement time with a {\Tm} exposure equal to \unit[$19.2$]{g$\cdot$day}.
From the data acquired, we set competitive limits on the axion coupling constants to electrons and photons.

The technology presented in this paper allows for a straightforward scaling of the experiment which would enable a drastic increase of the collected exposure.
Furthermore, the reduction of the background rate in the region of interest would translate into a considerable improvement on the sensitivity to solar axion absorption.
Since the background rate reduction can be effectively achieved with a dedicated underground experiment, we are confident in an improvement of the presented results in the near future.

\begin{acknowledgement}
This work was supported by Russian Foundation for Basic Research (grant numbers \#19-02-00097 and \#20-02-00571).
The authors also express gratitude to the support from NRC ``Kurchatov Institute'' provided by administrative order \#1808 of 14.08.2019.
\end{acknowledgement}

\bibliographystyle{h-physrev}
\bibliography{references.bib}

\end{document}